\documentclass[12pt,preprint]{aastex}

\begin{document}

\title{Red Giant stars
in the Large Magellanic Cloud clusters}

\author{Alessio Mucciarelli}
\affil{Dipartimento di Astronomia, Universit\`a 
degli Studi di Bologna, Via Ranzani, 1 - 40127
Bologna, ITALY}
\email{alessio.mucciarelli@studio.unibo.it}
\author{ Livia Origlia}
\affil{INAF - Osservatorio Astronomico di Bologna, Via Ranzani, 1 - 40127
Bologna, ITALY}
\email{livia.origlia@bo.astro.it}
\author{Francesco R. Ferraro}
\affil{Dipartimento di Astronomia, Universit\`a 
degli Studi di Bologna, Via Ranzani, 1 - 40127
Bologna, ITALY}
\email{francesco.ferraro3@unibo.it}
\author{Claudia Maraston }
\affil{University of Oxford, Denys Wilkinson Building, Keble Road, Oxford, 
OX13RH, United Kingdom}
\email{maraston@astro.ox.ac.uk}
\and
\author{Vincenzo Testa}
\affil{INAF - Osservatorio Astronomico di Roma, Via Frascati, 33 - 00040
Monteporzio Catone, ITALY}
\email{testa@mporzio.astro.it}

\altaffiltext{1}{Based on observations collected at the European Southern
Observatory, La Silla, Chile, using SOFI at the 3.5m NTT, within the observing
programs 64.N-0038 and 68.D-0287}

\begin{abstract}
We present deep J,H,Ks photometry and accurate Color Magnitude Diagrams 
down to $\rm K\approx 18.5$, for a sample of 13 globular
clusters in the Large Magellanic Cloud. This data set 
combined with the previous sample of 6 clusters published by our group
gives the opportunity to study the
properties of  giant stars in clusters with different
ages (ranging from $\sim 80~Myr$ up to $3.5~Gyr$).  
Quantitative estimates of  star population ratios (by
number and luminosity)  in the Asymptotic Giant
Branch, the  Red Giant Branch and the He-clump, have been
obtained and compared with theoretical models  in the framework 
of probing the so-called phase transitions. 
The AGB contribution to the total luminosity 
starts to be significant 
at $\approx 200~Myr$ and reaches its
maximum at $5-600~Myr$,
when the RGB Phase Transition is starting. 
At $\approx 900~Myr$ 
the full developing of an extended
and well populated  RGB has been completed.
Both the occurrence of the AGB and RGB Phase Transitions  
are sharp events, lasting a few hundreds $Myr$ only.
These empirical results agree very well with the theoretical predictions of simple
stellar population models based on canonical tracks and the
fuel-consumption approach.

\end{abstract}


\section{Introduction}
\label{intro}

The red giant star populations play an important r\^ole in
the spectral evolution of  a simple stellar population
(SSP), dominating the bolometric luminosity of a SSP  after
a few hundred million years. Two special events (the
so-called phase transitions,  hereafter Ph-T) 
due to the sudden  appearance of stars with C-O and
He degenerate cores mark the spectral evolution of a SSP. 
The first event corresponds to the appearance
of  bright Asymptotic Giant Branch (AGB) stars; the second
correspond to the full development of the Red Giant Branch (RGB).  
Several theoretical models \citep{rb86,bch93,m98} investigated 
the impact of the AGB and RGB
stars on the total  light of a SSP. The AGB Ph-T is
predicted to occur after $\approx10^{8}$ years and the  
RGB Ph-T  after $\approx6\cdot10^{8}$ years. 
These events must be empirically calibrated by using suitable
templates of  SSPs, in order to use them for predicting the
age of more complex stellar systems  like galaxies.   The
Large Magellanic Cloud (LMC) globular cluster system
represents a gold-mine  to study the AGB and RGB properties
with varying the age and  chemical composition of the
stellar population.  Indeed, these clusters cover a wide
range in age  \citep{swb80, ef85, girardi95, broc01},  in metallicity
\citep{sp89,ols91,hill00} and in 
integrated  colors \citep{vdb81, perss83},
providing  the ideal template to investigate the spectral 
behavior of a SSP.

The AGB and RGB phases are intrinsically difficult to
sample with the necessary  statistical significance,
because of their relatively short lifetimes. Several
searches investigated the ratio between C and M stars and
its correlation with metallicity in the LMC field
\citep{blanco80, cioni03, cioni05},  finding more C stars in
lower metallicity environments. \citet[][ hereafter FMB90]{fmb90} 
presented a detailed study of C and M-type AGB and
their contribution to the cluster total light    as a
function of \citet[][ hereafter SWB]{swb80} classes.  FMB90
assumed that the separation between AGB and RGB is
positioned at $M_{bol}$=-3.6 and stars brighter than that
value are on the AGB.  They found C-stars only in the
intermediate age SWB IV-VI clusters. The SWB VII clusters
are too old, and their AGB stars have too low mass stellar progenitors for 
evolving as C-stars, while SWB I-III clusters are still too young
to have stars evolving along the AGB. In these objects the
brightest red stars are M supergiants.  For intermediate
age SWB V-VI clusters 40\% of the bolometric luminosity is
contributed by AGB stars. In particular C-stars,
when present, account for 50-100 \% of the bolometric
luminosity of the AGB.\\
The observational results by FMB90 have been
used by \citet{m98,m05} to calibrate the energetics and
compositon of the TP-AGB phase to be inserted in stellar population
models. It is therefore very interesting to check those observational
results with new, accurate, data.

The first observational evidence of the RGB
Ph-T in stellar clusters has been
presented by \citet[][ hereafter F95]{f95} who analyzed
the red giant contents of 12 intermediate age LMC
cluster.  This survey was performed 
with the first generation of IR imagers, and sampled
only  the brightest portion (K$<$14.3) of the RGB. 
Despite of its limited luminosity sampling,
this pioneering work 
demonstrated that near-IR observations are indeed crucial in order to
study the photometric properties of AGB and RGB
sequences, providing the highest sensitivity  to the
physical parameters of cool stars.

Recently, the superior performances of the the 1024x1024
IR array detector of SOFI at the ESO/NTT, allowed to
measure the entire extension of the RGB for a complete sample of  LMC
clusters spanning a wide range of ages.  
\citet[][ hereafter F04]{f04}   
presented deep J,H,Ks photometry and accurate Color
Magnitude Diagrams (CMD) down to $\rm K\approx 18.5$, i.e.
$\approx 1.5$~mag  below the red He-clump,  for six
intermediate age clusters (namely NGC~1987, NGC~2108,
NGC~2190, NGC~2209, NGC~2231, NGC~2249).  A
quantitative estimate of the population ratios (by number
and luminosity) between Red Giant Branch and He-clump stars
for each target cluster, and comparison with  theoretical
predictions suggest that the full development of
the RGB should  occur at $t\approx 700~Myr$
and be a  relatively fast event ($\delta t\approx
300~Myr$).

The present paper shows the results for 13 additional LMC
clusters, extending the discussion  about the RGB Ph-T
presented in F04 and studyng the contribution to the total
luminosity of the AGB. The paper is organized as follows:
Sect.~2 describes
the observations and photometric analysis, Sect.~3 the
near-IR CMDs. Sect.~4 present the basic assumptions on reddening and 
the adopted age calibration. Sect.~5 
discusses the procedure adopted
to derive complete and accurate star counts and luminosity
contribution. Sects.~6 and 7 discuss the phase transitions, 
while in Sect.~8 we draw our conclusions.

\section{Observations and photometric analysis}
\label{obs}

J,H,Ks images of 13 globular clusters in the LMC  (see
Table~\ref{tabpar}  have been obtained at ESO, La Silla, on December
28-30, 2001, with the near IR imager\-/\-spec\-trometer
SOFI \citep{mcl98}  mounted at the ESO 3.5m NTT. SOFI is
equipped with a  $1024\times1024$ Rockwell IR-array
detector. All the  observations presented here, have been
performed with a scale of $0.292''/pixel$, providing a
$\approx 5'\times 5'$ field of view, each frame. The
observations were obtained in good seeing conditions
($0.7''-0.8''$ on average). Total integration times of 2
min in J, 4 min in H and 8 min in Ks  split into sets of shorter
exposures  have been  secured, allowing to obtain
accurate  (S/N$\ge$30) photometry down to 
J$\approx$19 and H,Ks$\approx$18.5.

Clusters have been selected accordingly to the 
{\it s}-parameter,  defined by \citet{ef85} as a curvilinear
coordinate running along the mean locus defined by the (U-B) and (B-V) 
integrated colors of the LMC
clusters, and directly linked to the cluster age. 
The 6 clusters
presented by F04 have $s-$parameter ranging between 34 and 37. 
The database presented here include 7 additional clusters
with $s=23-31$ and 6 with
$s=37-45$. 
Table~\ref{tabpar} lists the basic properties of the clusters
discussed in the paper: {\it s} values 
\citep{ef88} ,  $\rm E(B-V)$  \citep{perss83}
and metallicity  \citep{sp89,ols91,hill00,dirsch00}.
Indicative [Fe/H] values vary by a factor of $\approx$10, between -1.2 and solar  
with an average value of $\approx-0.6\pm 0.1$,
but precise abundance determinations via high resolution spectroscopy 
still lack for the majority of the LMC clusters.

A control field a few arcmin away from each cluster center 
has been also observed using the same instrumental
configuration,  in order to construct median-averaged sky
frames and for decontamination purposes.  A large sample of
high S/N flat-fields in each filter have been acquired  by
using a halogen lamp, alternatively switched on and off. 
The final cluster and control field frames have been 
sky--subtracted and  flat-field corrected. For each cluster,
the dataset includes three J,H,Ks images,  centered on the
cluster and three corresponding frames of an adjacent
field. 

The photometric analysis was performed by using {\it daophot-II} 
\citep{stet87}.
For each observed field all the images in the
J,H,Ks filters  were carefully aligned and trimmed in order
to have three output images,  one per filter, slightly
smaller than the original ones but perfectly registered. 
Then, a {\it daophot-II} Point Spread Function (PSF) 
fitting run was applied to each image. The output catalog
with the instrumental magnitudes has been checked for any
spurious detection or missing object  (typically 3--4 stars
at most) which have been included in the catalog  by hand.

Each catalog typically contains 1000-1200 stars, with the
exception of  3 poorly populated clusters counting less
than 700 stars (NGC 1651, 2162 and 2173)  and 2 rich
clusters with more than 1900 stars (NGC 1783, 1978). The
instrumental magnitudes have been transformed into the
Two-Micron  All-Sky Survey (2MASS) photometric system, by
using the large number of stars  (typically a few hundred)
in common. The overall dispersion of these transformations is $\sigma\le0.01$~mag 
in all the three filters. 
For sake of homogeneity, the output catalogs of F04
(calibrated on the \citet{persson98} photometric system)
have been also transformed  into the 2MASS one, although
the difference between the two calibration turns out to be 
negligible ($\approx 0.01$ mag).

The calibrated photometric catalogs in each filter were
finally matched  and merged together in a global catalog,
using  the CataXcorr and Catacomb softwares developed at
the Bologna Observatory  for an optimized cross-correlation. 

\section{Color--Magnitude Diagrams}
\label{cmd}

\begin{figure}
\plotone{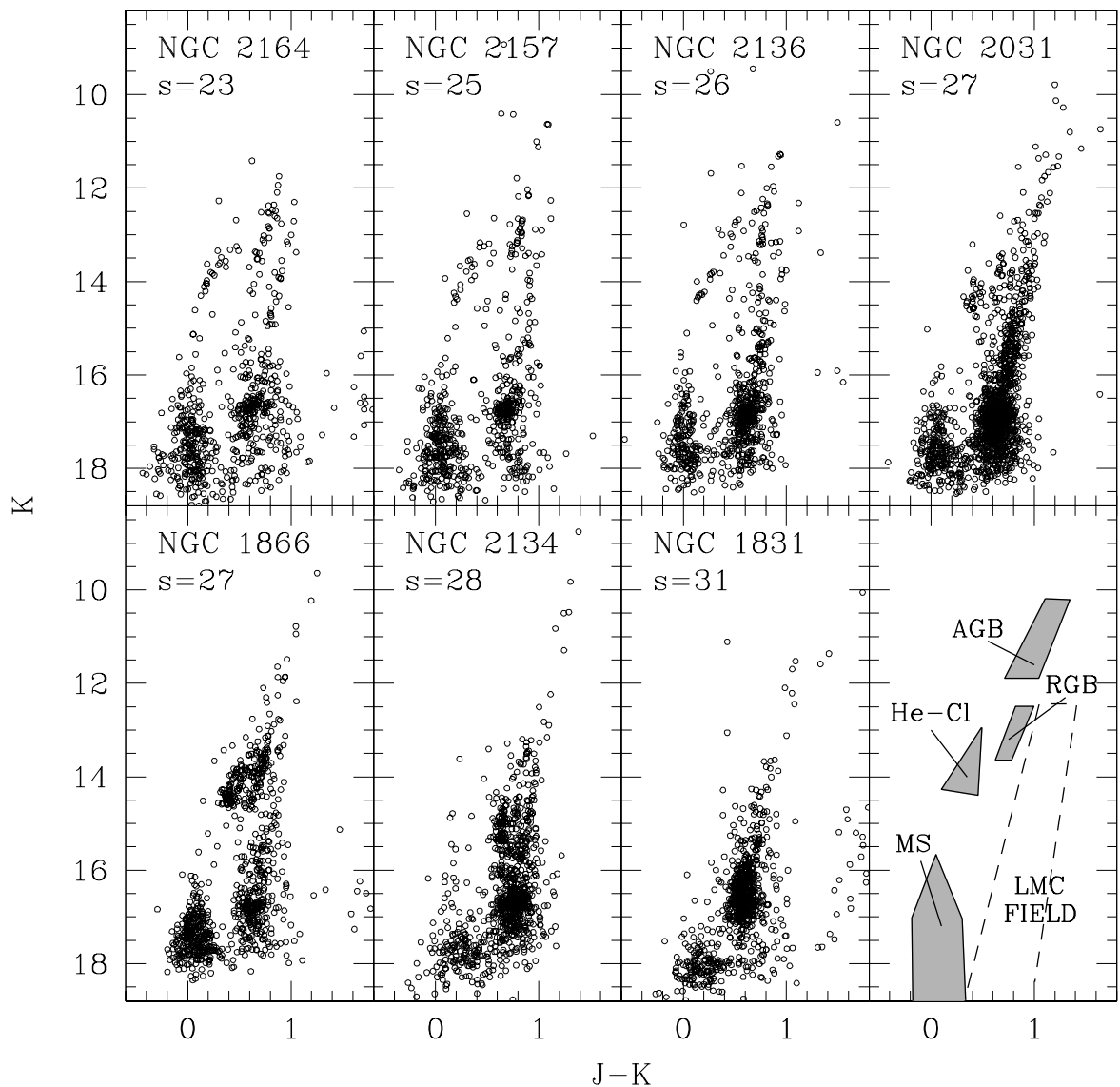}
\caption{Observed (K, J--K) CMDs of the 7 observed LMC clusters with $s=23-31$.
In the last panel a sketch of the CMD loci dominated by the 
cluster (grey regions) and the
LMC field (dashed box) populations are also shown for sake of clarity.}
\label{cmdc1}
\end{figure}

\begin{figure}
\plotone{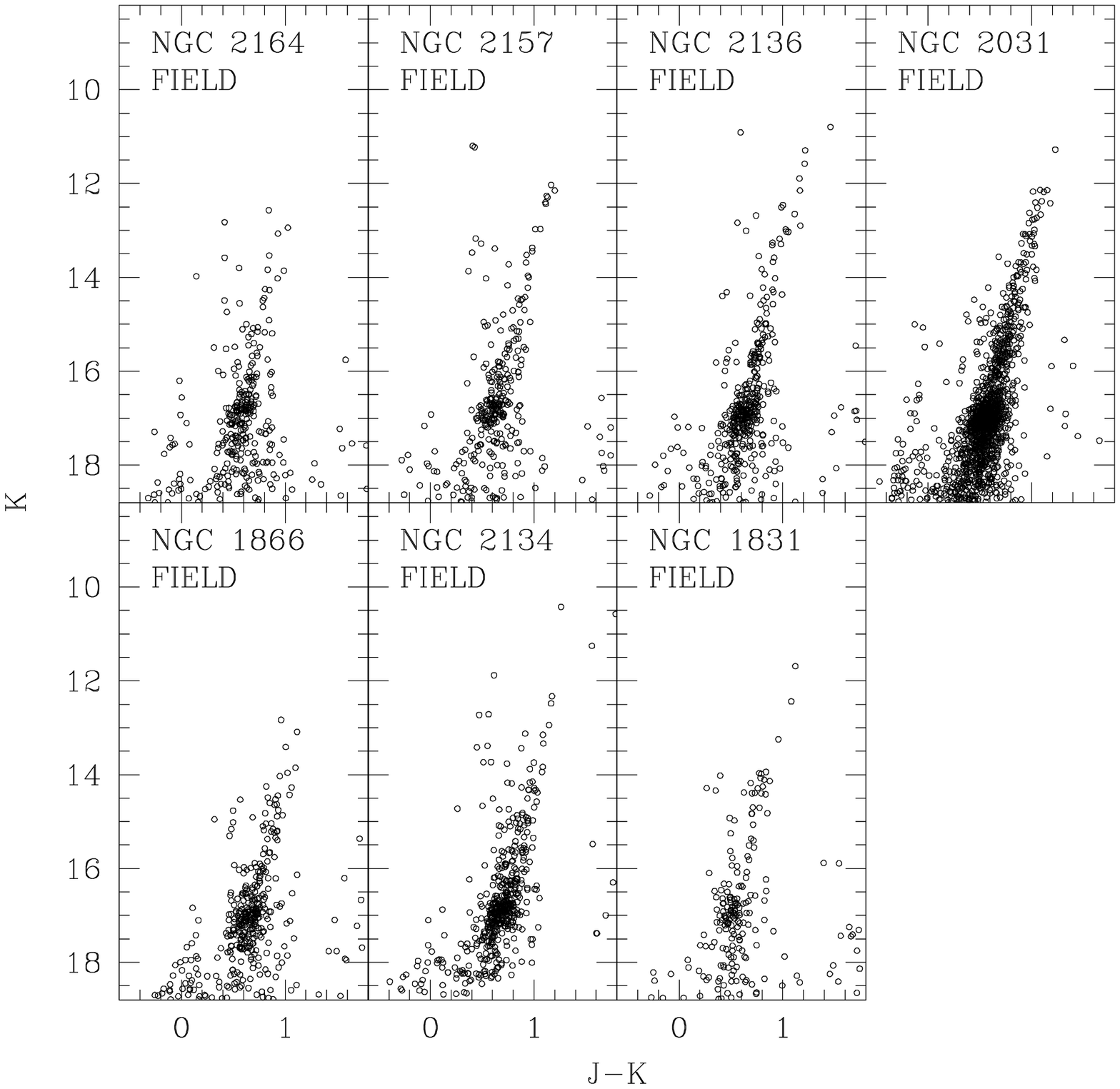}
\caption{Observed (K, J--K) CMDs of the fields adjacent to 
 the 7 observed LMC clusters with $s=23-31$.}
\label{cmdf1}
\end{figure}

\begin{figure}
\plotone{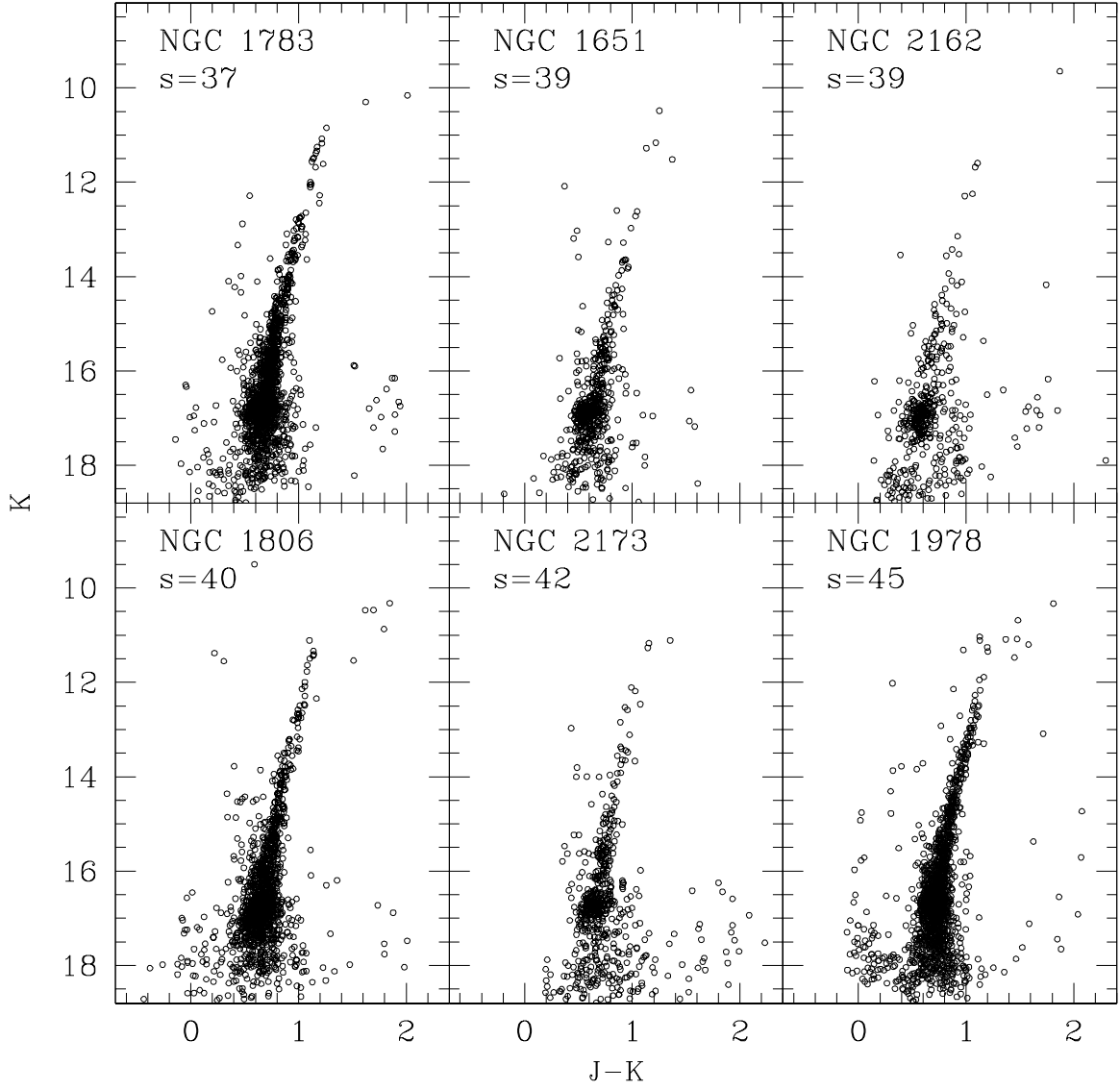}
\caption{Observed (K, J--K) CMDs of the 6 observed LMC clusters with $s=39-45$.}
\label{cmdc2}
\end{figure}

\begin{figure}
\plotone{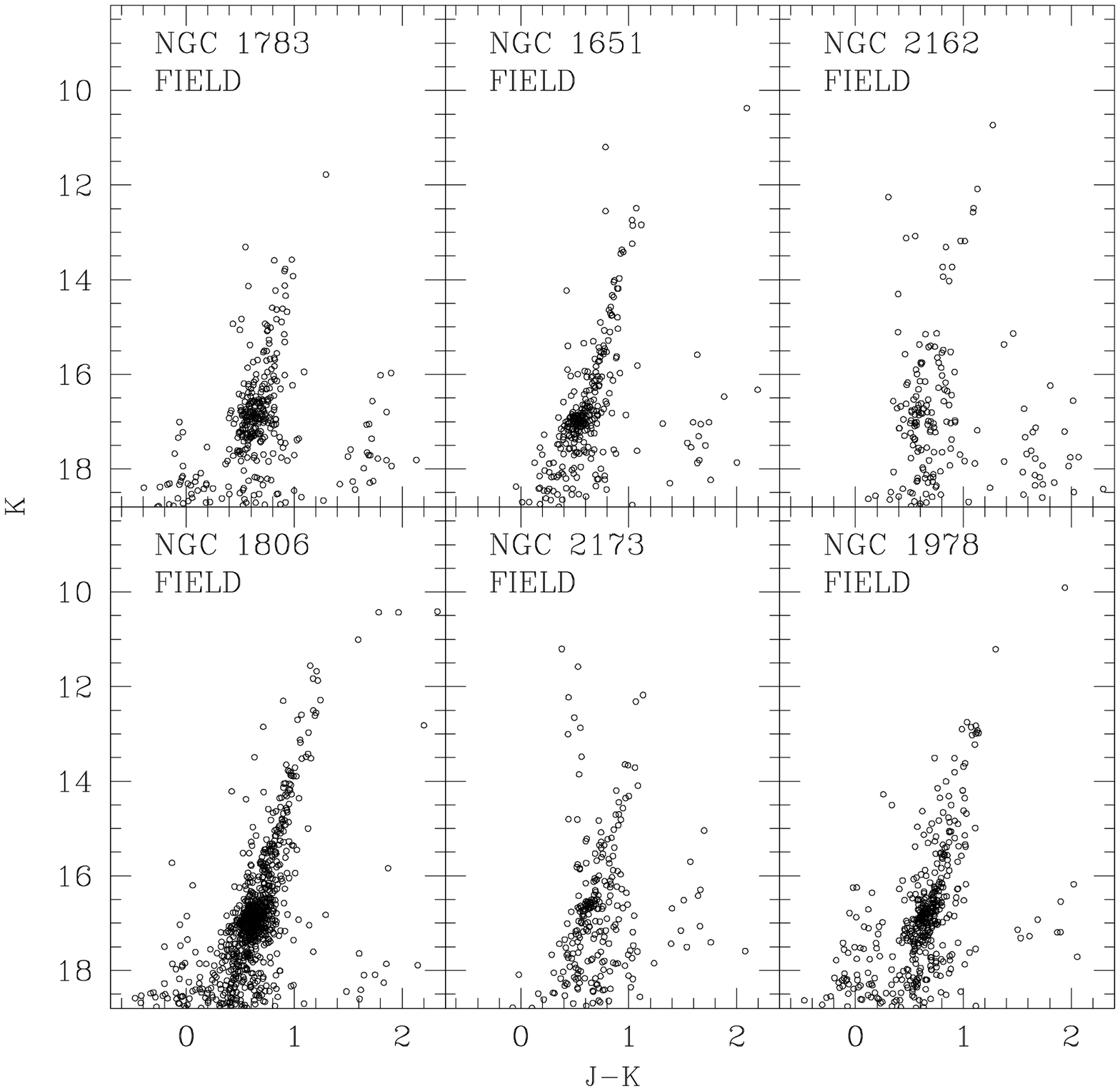}
\caption{Observed (K, J--K) CMDs of the fields adjacent to 
the 6 observed LMC clusters with $s=39-45$}
\label{cmdf2}
\end{figure}

Figs. 1 - 4 show the CMDs for the observed clusters  and
adjacent fields in the $\rm (K,(J-K))$ plane.  We have
divided the clusters into two groups, accordingly to their
s-parameter  and different CMD morphology. The first group
(see Fig.~\ref{cmdc1}) includes clusters with 
$s=23-31$;  the second group (see Fig.~\ref{cmdc2}) those with
$s=37-45$. Figs. \ref{cmdf1} and \ref{cmdf2}
show the CMDs of the corresponding adjacent fields. 
\footnote{It is
evident the extreme field contamination for NGC 2031, located in
the LMC Bar.}

The CMDs of the first group of clusters (with $s<31$)
appear quite complex, hence particular care has been 
devoted  to separate the cluster population from the 
LMC field. In order to help the reader
identify the two populations we have plotted in the last 
panel a sketch showing the mean location of cluster 
(grey regions) and field (dashed region) population in the 
CMD. The main properties of the CMDs shown in Fig.~ \ref{cmdc1}
can be summarized as follows:
  
{\it (i)~}Magnitudes as faint as K$\approx18.5$ have been measured.

{\it (ii)~}The brightest objects at $\rm K<13$ are likely AGB stars.

{\it (iii)~}A blue sequence is clearly visible 
at -0.3$\rm<(J-K)<0.3$ and K$>$15.5, 
corresponding to the brightest end of the cluster MS.

{\it (iv)~} Helium clump cluster stars
define a sequence at $K=13-14$ and
$(J-K)\sim 0.2$ in the youngest objects (namely
NGC2164, NGC2157, NGC2136). 
In older clusters (namely
NGC2031, NGC1866, NGC2134, NGC1831),
they define a 
clump at progressively lower magnitudes ($K=14-15$)
and redder colors ($\rm 0.4<(J-K)<0.6$).

{\it (v)~} The direct comparison of each panel in
Figs. \ref{cmdc1} and \ref{cmdf1} clearly shows the
significant contribution of the LMC field stars.
As schematically shown in the last panel of Fig.\ref{cmdc1}, 
most of the stars in the region 
$0.4<(J-K)<1$ and $K>12$ are indeed LMC field stars,
with a well defined RGB at 
$12<K<16$ and the He-clump at K$\approx$17. 

The observed CMDs for the second group of cluster show the
following characteristics: 

{\it (i)~}Magnitude limits down to K$\approx18.5$,
i.e. about 1.5~mag below the He-clump, which is
clearly visible as a clump of stars at $\rm K\approx 17$.

{\it (ii)~} A well-populated and extended RGB. 

{\it (iii)~}The brightest objects at $\rm K<12$ are likely AGB stars.

{\it (iv)~} Unlike the first group of clusters, in these
objects cluster and field  populations do overlap.

\section{Basic assumptions}

\subsection{Reddening} 

Correction for extinction is computed accordingly to the $\rm
E(B-V)$  values reported in Table~\ref{tabpar} and the
\citet{rl85} interstellar extinction law.  The infrared
dust maps by \citet{sfd98} in the direction of  the
observed clusters provide very similar  (on average within
$\pm$0.03~dex) $\rm E(B-V)$ corrections,  with the
exception of NGC 2031,  for which the discrepancy is about
0.1~dex. However, the overall impact of such a discrepancy
on the infrared magnitudes  is always small (well within
0.1~dex), hence reddening correction is not  a critical
issue in this context.

Absolute and bolometric magnitudes have been obtained by adopting a
distance modulus $\rm (m-M)_0=18.5$ \citep{vdb98,alv04} and 
suitable bolometric corrections by using the $\rm
(J-K)_0$ color and the empirical calibrations by
\citet{mon98}. In computing luminosities, we adopted 
$M^{Bol}_{\odot}=4.74$ and $M^{K}_{\odot}=3.41$ for the Sun. 
In the following
all the derived luminosities are expressed  in unit of  $10^4 L_{\odot}$.

\subsection{Age calibration}
\label{age}
 
A suitable calibration of the LMC cluster age is still a
major concern since  homogeneous determinations based on
the MS TO, for a significant
number of clusters, are not available yet. Following F04, here we
use the $s-$parameter 
(see Sect.\ref{obs}). Being a pure
empirical quantity,  
it needs to be calibrated with
age. As discussed in F04, the most used calibrations  by
\citet{ef88} based on canonical models and by 
\citet{girardi95} based on  overshooting models provide,
somewhat surprisingly, very similar ages (within 10-15 \%). 
Although a new calibration
of the {\it s}-parameter as a function of age (see F04)
is urged, in the following, as done in F04, we adopted the most 
recent one by  \citet{girardi95} and
based on the models with overshooting by Bertelli et al.
(1994): $$\log{t}=6.227+0.0733\cdot{s}.$$  We assume a
conservative error of $\delta s$=$\pm$1, which translates
into a  $\approx$ 20\% age uncertainty. 

Table~\ref{tabpar} lists the $s$ value for the entire sample of
19 clusters (13 presented here and 6 discussed  in F04,
respectively) and their age
derived  from the quoted calibration. As
can be seen the entire sample cover a large range of ages
from $80~Myr$ to $3.5~Gyr$.

\section{Star counts and integrated luminosities} 
\label{count}
 
A quantitative analysis of the AGB and RGB populations
(by number and  luminosity) is crucial
to empirically calibrate the relative  lifetimes and to
quantitatively evaluate the impact of each
evolutionary stage  on the
total luminosity of a SSP.
In order to obtain reliable stellar
counts and  luminosities in each branch,
we proceeded as follows:
{\it (1)} stars in each evolutionary stage
have been identified on the basis of suitable selection boxes 
as defined in the
CMDs (as shown in  Fig.~\ref{cumagb} and Fig.~\ref{boxrgb});
{\it (2)}  
each sample of stars  has been  corrected for 
incompleteness, following the standard {\it artificial star}
technique (see below); 
{\it (3)} the contamination from foreground/background stars
in each population has been evaluated and statistically
subtracted to the observed samples.

While the definition of the selection boxes ({\it step (1)})
for the AGB and RGB populations are described in Sects.~\ref{agbpht} 
and \ref{rgbpht}, respectively, 
in the following we
briefly discuss  the procedure adopted to
perform {\it step (2)} and {\it step (3)}.

\subsection{Completeness and field decontamination}
\label{comp}
 
\begin{figure}
\plotone{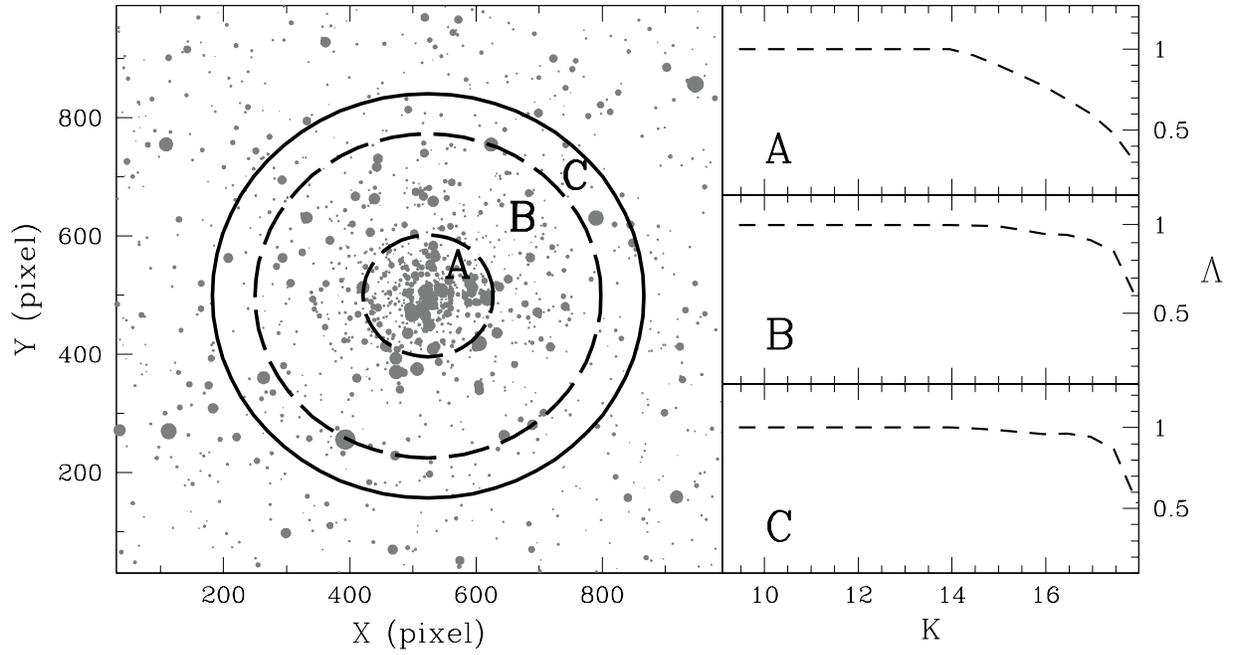}
\caption{An example of cluster radial mapping.
{\it Left panel}: the cluster frame is divided into 3 concentric annuli, to account 
for different crowding conditions. 
{\it Right panel}: completeness curve for each radial sub-region, as labeled 
in the left panel.}
\label{map}
\end{figure}

The degree of completeness can be quantified by  using the
widely-used artificial star technique  \citep[see the discussion
present in][]{mateo88}.  For each cluster we have
derived the RGB fiducial line and then a population of
artificial stars, having magnitudes, colors, and luminosity functions 
resembling the observed distributions was  generated and
added to the original images.  Since crowding effects are
more severe in the central regions, the frame area
sampling the cluster  has been
divided in three concentric regions ({\it Region A, B and
C}, see  left panel of Fig.~\ref{map}) and the completeness
has been estimated independently in each of them. The maximum
spatial extension of each cluster  has been estimated from
the cluster radial density profile.  
A total of $\approx$200,000
artificial stars   have been simulated  in each cluster  in
about 1000 simulation runs. Indeed, in order to not alter the crowding
conditions, only 100--200 stars have been simulated in each run.
The fraction of recovered objects in each magnitude
interval was estimated  as
$\Lambda=\frac{N_{rec}}{N_{sim}}$,  and a suitable 
completeness curve was obtained in each of the A,B,C regions
(see right panels of  Fig.~\ref{map}).  

Star counts in each radial region
have been finally corrected for incompleteness, by
dividing each observed distribution by the corresponding
$\Lambda$ factors\footnote{Note that the number
of stars lost for incompleteness  ($n_{comp}$) in each bin of magnitude
is $$n_{comp}= n_{obs} (1/\Lambda -1)$$
where $n_{obs}$ is the number of stars observed in that bin.}.
The total number of stars   
has been finally obtained
by summing the completeness-corrected number of stars of 
the A,B,C regions. 

As already discussed in previous papers,
the artificial star technique provides only a first-order 
correction. In fact, the
observed  distribution is, in principle, distorted because of
two main phenomena:  the loss of faint stars due to
incompleteness and an excess of bright stars  due to
possible blending effects of two or more faint stars into a
brighter one.  Only the first effect is taken into account
by the artificial star  simulation. Blending from faint MS
stars is a more complicated effect to simulate,  however in
the near IR it is negligible  
\citep[see also discussion in][]{testa99}.

Another important effect which needs to be investigated, is
the degree  of contamination of the selected samples by the
foreground/background stars.  In this paper we have applied
a statistical decontamination technique,  using a control
field adjacent to the clusters.
The total number of stars observed in each evolutionary
sequence  (AGB, RGB and He-clump) has been counted
accordingly to the {\it selection boxes}  both in the 
cluster (see Figs.~\ref{cmdc1} and ~\ref{cmdc2}) and field  
(see Figs.~\ref{cmdf1} and \ref{cmdf2}) CMDs, and 
corrected for incompleteness (see above).  
The star counts in the field population have been
scaled to take into account the different surveyed area, 
and their contribution have been subtracted from the
cluster population. 

In summary, 
for each radial region, each selection box 
corresponding to each evolutionary stage has
been divided in bins of magnitude (typically 0.2 mag
wide). Then, the "corrected" number of stars in each bin
has been computed as follows:
$$ n_{corr} = n_{obs} + n_{comp} - n_{f}$$
where $n_{obs}$  is the  number of stars observed in that bin,
$n_{comp}$ is the number of stars lost for incompleteness,
$n_{f}$  is the expected  number of  
field stars. 

Analogously, the  
total luminosity of each evolutionary stage can be computed
accordingly to the following relation:
$$ L_{corr} = (\sum^n_{i=1} L^{obs}_{i}) + (n_{comp}\times L_{eq}) -
(n_{f} \times L_{eq})$$
where the term $\sum^n_{i=1} L^{obs}_{i}$ is the total luminosity of the
stars observed in a given bin, $n_{comp}$ is the number of
stars lost for incompleteness, $n_{f}$  is the expected 
number of field stars, and $L_{eq}$ is the equivalent
luminosity of that bin, that is the luminosity of a star with
magnitude equal to the mean value of the bin.

Finally, star counts and total luminosity of each
evolutionary stage have been obtained by summing the  
contribution of all the bins in each selection
box.

\subsection{Integrated magnitudes}
 
In order to properly perform cluster to cluster
comparisons,  one needs to take into account the size of
the total cluster population. Hence both star counts and
luminosities needs to be normalized to a reference
population or to the cluster integrated luminosity. In
previous papers we followed both the approaches (in F95 and
F04 the RGB populations were normalized to the cluster
integrated luminosity taken from the literature,  and  to
the He-clump population, respectively).

The large FoV of SOFI offer the opportunity to
independently  determine the  near-infrared integrated
magnitudes for the program clusters. In doing this, we
adopted a simple approach, by performing aperture
photometry  over the entire cluster extension
(typically $90''$).
In order to correct for the 
field contamination, an 
equivalent aperture photometry has been also performed on each
control field and  the resulting luminosity has been 
subtracted from the cluster value. 

The
cluster center has been computed by applying a standard
technique -- \citet[see for example][]{calzetti93} -- which uses
the knowledge of the position of individual stars in the
innermost region of the cluster, allowing a high precision
determination of the center of gravity. 
Hence, by applying the procedure described in \citet{mon95}
we computed $C_{\rm grav}$ by simply averaging
the $\alpha$ and $\delta$ coordinates of stars lying within
a fixed radius from a first-guess center estimated by eye. 
The barycenter of the stars is then derived iteratively (see also
\citet{fer03}). The center of gravity 
($C_{\rm grav}$)  of the programme clusters 
are listed in Table~\ref{tabpar}.
Our new estimates turn out to be reasonably consistent 
(within $\sim 10"$) 
with available determinations (as those in the 
SIMBAD astronomical database by the CDS, Strasbourg). The typical
$1\sigma$ uncertainty of our estimates is $\sim 5$
pixels corresponding to $1.5\farcs$ in both $\alpha_{\rm
J2000}$ and $\delta_{{\rm J2000}}$.
The position of the center of 2 clusters, 
namely NGC2136 and NGC2173,
appear to be significantly 
(up to 2 arcmin) different from the SIMBAD coordinates. 

The case of NGC~2136 deserves an additional comment. 
Indeed, a small {\it twin } cluster, namely NGC~2137, is
present at an angular distance of $1.34'$  \citep{hilk95}.
Since its integrated luminosity, although significantly fainter,
is contaminating the aperture photometry of NGC~2136, it 
has been properly subtracted.  

Integrated K magnitude, colors  and derived bolometric luminosities in
the K band and in bolometric for the entire sample of 19
clusters are listed in Table~\ref{tabmag}.

\subsection{Error budget}
\label{err}

Formal errors are directly estimated from
the photometric samples, by assuming that 
star counts follow the Poisson statistics. 
The 
errorbars for the various population ratios (by number and/or by luminosity) 
have been computed accordingly to 
the following formula 
$$\sigma_{R}=\frac{\sqrt{R^{2}\cdot\sigma_{D}^{2}+\sigma_{N}^{2}}}{D}$$ 
with $R=N/D$, $N$ being the numerator and $D$ the denominator of the ratio.
  
In the computation of the population ratios different error
sources are at work,  depending on the observable.
\begin{itemize} 

\item total cluster luminosity: the main source of 
uncertainty  in this case  is the positioning of the
cluster center. We estimate that an off-centering of 5 pixels corresponds to
a 5\% variation in luminosity. An additional uncertainty  
of $\approx 10\%$  has been considered in the 
computation of bolometric luminosities, in order
to take into account the
uncertainty in the bolometric corrections.

\item AGB luminosity: a conservative 
uncertainty of 0.2~mag in setting the faint 
end of the AGB luminosity distribution, implies a $\approx$5\%
variation in the total AGB  luminosity. However, for this
observable the major source of uncertainty is the
random error associated to the number of detected AGB
stars (in the Poisson regime $\sigma \propto \sqrt N_{AGB}$), 
which can suffer large fluctuations  due to the
small number statistics.  
On average, the overall $\sigma_{R}$ associated to the $\frac{L_{AGB}^{K}}{L_{TOT}^{K}}$ 
ratio turns out to be 30\%.   

\item number and luminosity of C-stars: as in the
case of AGB stars, these observables and  their associated
errors suffer large fluctuations due to  the small number
statistics. 
On average, the overall $\sigma_{R}$ associated to the $\frac{L_{C-star}^{K}}{L_{TOT}^{K}}$ 
ratio turns out to be $\approx$50\%.

\item RGB luminosity: in the older 
clusters (see Fig.~\ref{cmdc2}) the RGB is well populated, 
hence the estimated luminosity 
is much less affected by statistical fluctuations or by 
the selection box definition. 
On average, we estimate a $\sigma_{R}\approx$20\% for the RGB population ratios.    
\end{itemize}

\section{The AGB and C-rich star contribution to the cluster light}
\label{agbpht}

Theoretical models \citep{rb86,m98,m05} predict that the
most  important contributors to the integrated cluster
light between $10^{8}$ and $10^{9}$ yrs  are AGB stars. The
AGB population includes both O-rich (M-type) and 
C-rich stars.  During the
thermal pulsing phase (hereafter TP-AGB) an AGB becomes C-rich  if it
undergoes the third dredge-up mixing process   \citep[see e.g.][]{ir83}. 
The presence of C-stars in stellar clusters
depends on  their age and metallicity \citep{rv81}. 

In intermediate age clusters the bulk of the AGB population
is more luminous  than the RGB Tip,  and a minor overlap does
exist between the faintest end of the  AGB and the
brightest portion of the RGB. Here we use our data set in
order to investigate the contribution to the cluster
luminosity of the brightest portion of the AGB populations
as a function of the cluster age. In order to consider the
entire database that spans a large range of ages 
(from $80~Myr$ to $3.5~Gyr$, see Table~\ref{tabpar}), in the following we consider 
only  AGB stars brighter than K$\approx 12.3$, which
represents the RGB Tip level for the oldest clusters in our
sample (see F04).

\begin{figure}
\plotone{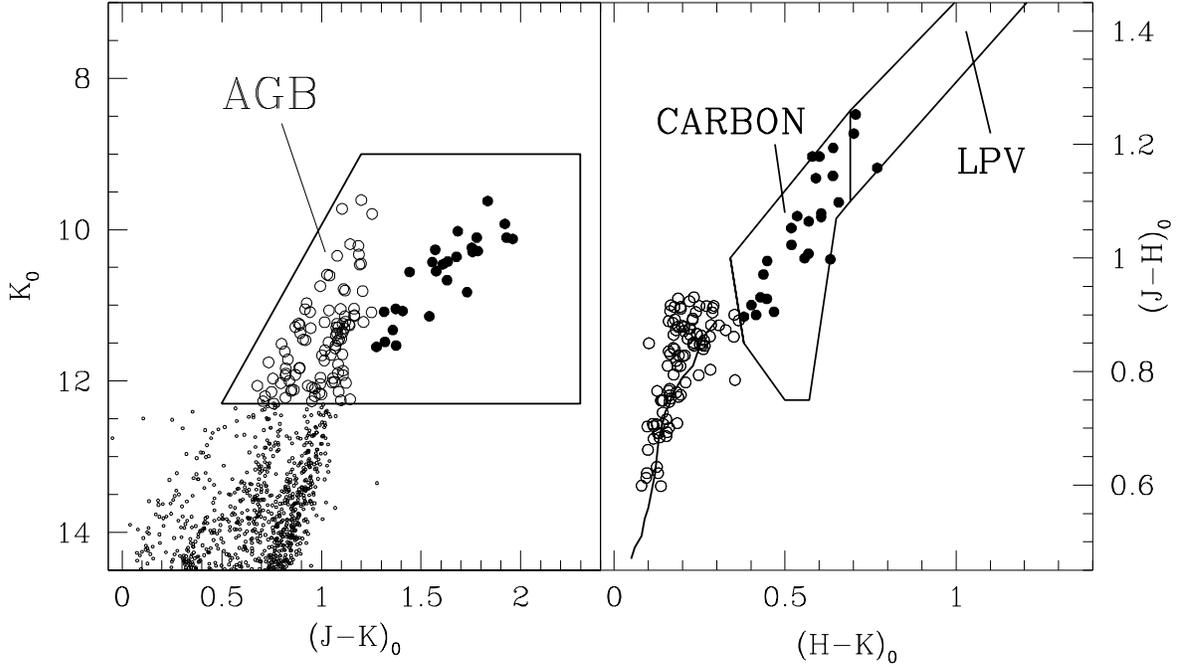}
\caption{{\it Left panel}: cumulative, de-reddened $\rm K_0,(J-K)_0$ CMD for the
entire cluster sample. The selection box adopted to isolate the  AGB population
({\it large circles}) is shown. 
{\it Right panel}: de-reddened color-color $\rm(J-H)_0,(H-K)_0$ diagram 
of the AGB stars. 
In both panels {\it open circles} are O-rich AGB,  {\it filled circles} are C-stars. 
{\it Solid line} boxes to distinguish C-stars and Long Period Variables (LPV) are from 
\citet{bb88}, \citep[see also][]{f95}. The mean locus for K giant stars ({\it solid line}) 
is from \citet{fpam78}.}
\label{cumagb}
\end{figure}

The left panel of Fig.~\ref{cumagb} shows the  brightest
portion of  the $\rm K_0,(J-K)_0$ cumulative CMD, where all
the stars detected in the 19  surveyed clusters are
plotted. The selection box adopted to sample the bright AGB
population is over-plotted to the diagram. The right
panel of Fig.~\ref{cumagb} shows the cumulative
$\rm (J-H)_0,(H-K)_0$ color-color diagram for the selected
AGB stars.   This diagram is especially suitable to
isolate C-stars, since they are significantly redder than  
O-rich stars \citep[see also][]{cioni05}. As
shown in Fig.~\ref{cumagb}  a population of 26 candidate
C-stars (plotted as {\it filled circles}) 
has been  identified on the basis of their
extremely red colors in the 19 surveyed clusters. 

 Note also that the artificial star experiments
 demonstrated that stars lying in the
brightest portion of the RGB can be safely recovered (with an overall
photometric uncertainty of $\sim0.03$ mag) even in the innermost region of the clusters,
excluding the possibility that blending of RGB stars could produce spurious
bright objects lying within the AGB selection box.

We used both the cumulative color-magnitude and color-color diagrams in Fig.\ref{cumagb} 
to make a census (both
by number and luminosity)  
of the AGB stars brighter than the RGB Tip as well as of C-stars in
each cluster.  Although the number and the luminosity of
AGB stars are subjected to large  fluctuations due to the
small-number statistics,  we still performed a statistical
decontamination, following the procedure described in 
Sect.~\ref{comp}.

The number of AGB stars counted in each clusters and the
number adopted after the field de-contamination  are listed in
Table~\ref{tabagb}.  Once  the accurate census of  the AGB population
(by number and luminosity) is available for all the
sampled  clusters, a number of suitable  diagnostics tools
can be used in order to study the AGB properties  
as a function of the age.

The upper panel of Fig.~\ref{agbkbol} shows the ratio between the AGB and the 
cluster integrated K-band luminosity,  as a function of the cluster
age. 
It is remarkable the rapid increase (up to a factor 2) 
of the AGB luminosity at  $\approx 200~Myr$ which reaches
its  maximum contribution in 
the 300-700~Myr range followed by a rapid decrease.    
Note that in 2 clusters (namely NGC2108 and
NGC1987 at $s=35-36$ corresponding to $t\sim600-700 Myr$)
the brightest portion of the AGB population account
for $\approx 90\%$ of the total cluster luminosity.
These results are in good agreement with  F95, who found
that the  maximum contribution of the AGB to the cluster
light occurs at   $s=35$, corresponding to an age
of $\approx 600~Myr$.

The lower panel of Fig.~\ref{agbkbol} shows  
the same ratio as in the upper panel but
with the clusters grouped in five age bins accordingly to 
their s-parameter, namely s=23-26, 27-31, 35-36, 
37-39 and 40-45, respectively. 
For each bin we computed the weighted mean and the corresponding standard deviation. 
Theoretical predictions from \citet{m98,m05} for [Z/H]=-0.33
 are also plotted.
In these models the TP-AGB energetics was calibrated with 
previous (FM90, F95) intermediate age MC cluster data (see 
\citet{m98} for full 
details).
Our new observations nicely confirm the modelling and those 
early results.
Old canonical models of stellar evolution
\citep{rb86} were dating the occurrence of the AGB Ph-T by
\citet{rb86} at significantly earlier epochs
($\approx10^{7}~yrs$) of the stellar lifetime with respect to
the new models.  This discrepancy is due to a different
treatment for the TP-AGB stars that experiencing the
envelope burning process \citep{rv81,bs91}, as
widely discussed in \citet{m98}.

As a further evidence, Fig.~\ref{carbon} shows the ratio
between the number of C-stars and total bolometric
luminosity and the ratio between the K luminosity of
C-stars and the total  K-band cluster luminosity, as a function of
the cluster age.  The number of C-stars in each cluster
is listed in Table~\ref{tabagb}.   
We assumed all these stars being
cluster members. Indeed, 
we estimate that the probability to find a field C-star
within the  sampled cluster area is $<$30\%. The
contribution of C-stars to the total cluster luminosity
as a function of the cluster age,  closely follows the one
shown by the entire AGB population  (see Fig.~\ref{agbkbol})
and it turns out to be larger than 50\%  in the $700-1000~Myr$ age range.
In NGC~2190
(with $s=36$, hence $t=730~Myr$) the
C-stars accounts for 70\% of its total luminosity.
Previous works by FMB90 also found that the fraction of
luminosity from  bright AGB and C-stars is maximum for SWB
V clusters (i.e. $s$ between 35 and 40).   

It is worth noticing the case of NGC~2249, whose age ($log~t=8.72$) 
corresponds to the epoch when the AGB contribution is expected 
to reach its maximum.
Conversely, both Figs.~\ref{agbkbol} and \ref{carbon} shows that  
NGG~2249 (marked with an open circle) has a very low AGB luminosity 
for its age.  Indeed no C-stars and only 1 AGB have been detected 
in this cluster. 
On the other hand, NGC2249 is the least luminous cluster in our sample
($\rm L_K\approx 5 \times 10^4 L_{\odot}$), hence the
fastest evolutionary stages (as the 
AGB) are expected to be intrinsecally poorly populated in its CMD.

\section{The RGB Ph-T}
\label{rgbpht}

For the analysis of the RGB Ph-T we have considered the 
clusters in our data-base with $s>34$
(the 6 oldest clusters in the sample presented here and
the 6  discussed in F04). Obviously the 7  
youngest clusters in our sample (Fig.~\ref{cmdc1}) are not 
considered in the following 
discussion because  they have not
developed a populous RGB yet. The  considered
sample  covers a wide  range of  ages
(from $500~Myr$  up to $3.5~Gyr$) significantly extending
the age range  covered by F04 and it allows to 
probe the entire developing of the RGB.

In order to compute the RGB population ratios we have
adopted the same procedure described in F04. Using the
cumulative $K_0$, $(J-K)_0$ CMD for the 12 clusters we
identified the mean loci of the upper RGB and the He-Clump
and defined  the corresponding selection boxes sampling
these  populations (see  F04). As discussed in
Sect.~\ref{agbpht}  the RGB Tip is expected to be at
$K_{0}\approx$12.3. As an example, Fig.~\ref{boxrgb} shows
the de-reddened CMD of NGC~1651 where the  two selection
boxes for the RGB and He-clump population, respectively, have been plotted.

Since the photometric errors can significantly broaden the faint sequences,
the size of the boxes
including the base of the RGB and the He-clump
has been conservatively assumed to be $\approx$ 5 times the photometric error at
that level of magnitude\footnote{Note that, since the boxes sample the bulk of the population along 
 each evolutionary stage, a slightly different assumption in the 
  selection box size has a negligible impact on the overall results.}.

Population counts and luminosities for stars in the RGB and
He-clump evolutionary stage have been obtained and
corrected for incompleteness and field contamination
accordingly with the procedure discussed in
Sect.~\ref{comp}. 
The results are listed in Table~\ref{tabrgb} and
plotted in Figs.~\ref{contrgb} and \ref{lumrgb}. Fig.~\ref{contrgb} 
shows the
behavior of the  number of RGB stars  normalized to the number
of He-Clump stars  as a  function of the cluster age. 
Fig.~\ref{lumrgb} shows the bolometric luminosities of RGB
stars normalized  to the bolometric luminosities of
He-Clump stars as a function of the cluster age (upper panel) and 
the bolometric luminosities of RGB
stars normalized  to the bolometric luminosities of 
entire cluster as a function of age (lower panel).

Note that, in a few high density clusters (see Table~\ref{tabrgb}) 
with severe crowding, 
completeness drops down to 60\% 
at the He-clump magnitude level 
in the innermost A region (see Fig.\ref{map}). 
Hence, in these clusters star counts and luminosities have
been computed only in the outer B and C regions.

As already discussed  in F04, at an age of $\approx 500~Myr$ 
the  rapid increase of the RGB population ratios   (by a factor of
$\approx$3 in number and $\approx$4 in luminosity)  in a
timescale as short as $\approx 400~Myr$ flags the occurring
of the RGB Ph-T. At the age of
$\approx 900~Myr$ a progressive flattening of the ratios
suggests that the full developing of an extended
and well populated  RGB   has occurred.
 The overall increase of the population ratios
between $\approx 500~Myr$ and $\approx 3.5~Gyr$  is a factor
$\approx$5 by number and $\approx$7 by luminosity.\\ 
Empirical data have been compared to theoretical
predictions. 
In Fig.~\ref{contrgb} and Fig.~\ref{lumrgb}, we 
over-plotted the predictions of
canonical models \citep{m98,m05} with $[Z/H]=-0.33$.
The models nicely agree with the observations over the 
entire range of considered ages, well describing the epoch,
the duration and the increasing contribution of the RGB
phase. 

It is worth noticing the population ratio excess (both by
number and by luminosity,  see Figs.~\ref{contrgb} and
\ref{lumrgb}) of NGC~1783, when compared with other
clusters with similar values of the $s-$parameter (i.e., NGC~2231). 
It is likely that this
cluster be older than suggested by the $s-$parameter, since 
its CMD (see Fig.~\ref{cmdc2}) shows a fully populated RGB,
typical of clusters with s-parameter$\ge$40. 
Indeed, its RGB morphology 
is more similar to that one of clusters such as NGC~1806, NGC~2173 and NGC~1978 
rather than that one of NGC~2231 (see Fig.~2 in F04). 
This evidence further supports the urgency of a new
homogeneous calibration of the age scale of LMC clusters.

\section{Conclusions}

We have used our extensive  near IR database of 19
LMC clusters spanning a wide range of ages (from $80~Myr$ to
$3.5~Gyr$) in order to investigate the developing  
in time of the AGB and RGB evolutionary stages.

\begin{figure}
\plotone{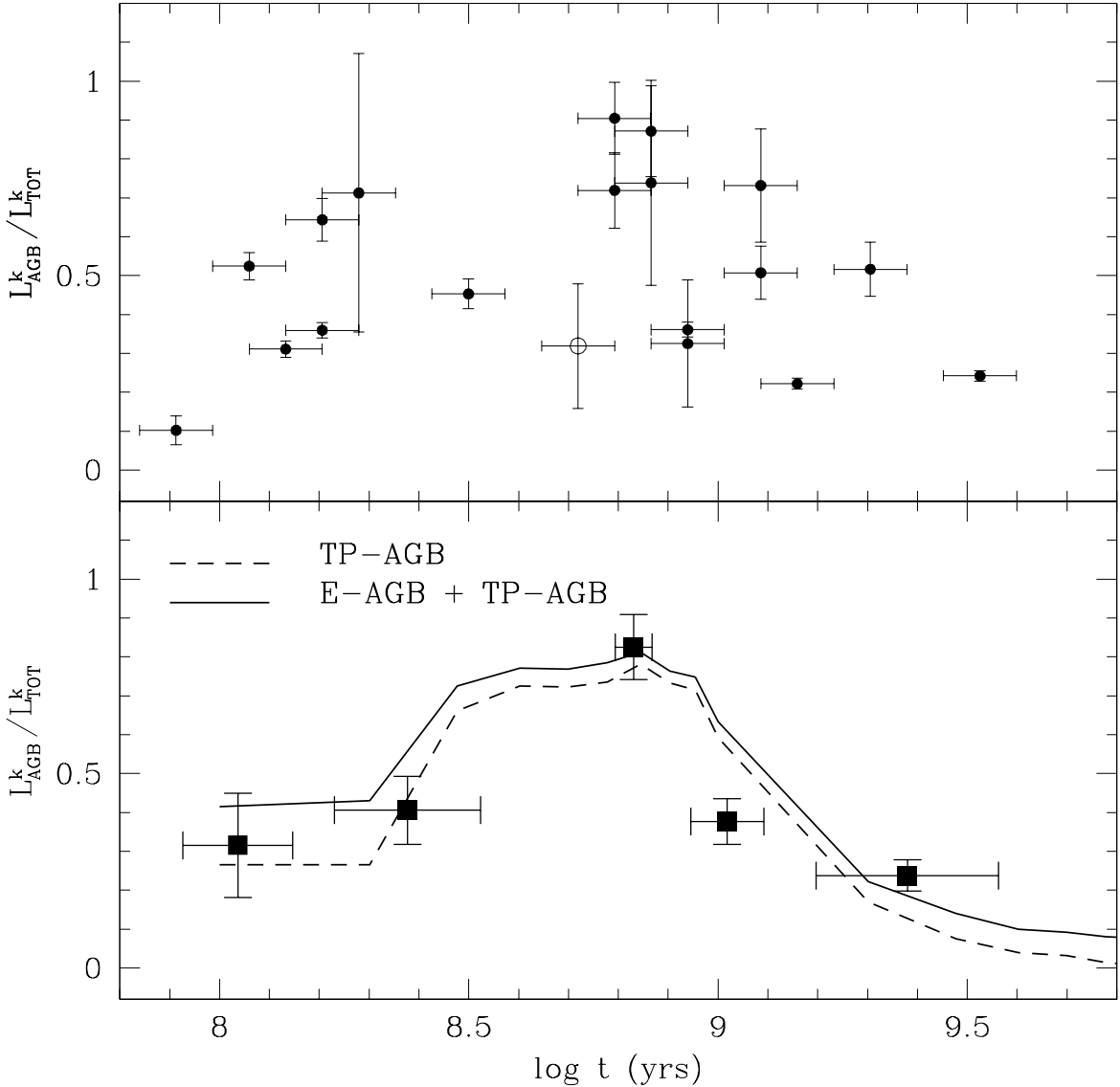}
\caption{Upper panel: observed AGB contribution 
to the total cluster K-band luminosity
as a function of age. 
The open circle marks the intrinsically poor populated cluster NGC~2249. 
Lower panel:  weighted mean and standard deviation of the same ratio 
with the clusters grouped in five age bins, 
accordingly with their s-parameter, namely s=23-26, 27-31, 35-36, 
37-39 and 40-45, respectively 
(the cluster  NGC~2249 has been excluded). 
Theoretical predictions for the 
temporal evolution of the entire AGB (E-AGB and TP-AGB, {\it solid line})
and for the dominant TP-AGB ({\it dashed line}) are overplotted. Both models
are computed  at [Z/H]$=-0.33$.}
\label{agbkbol}
\end{figure}

\begin{figure}
\plotone{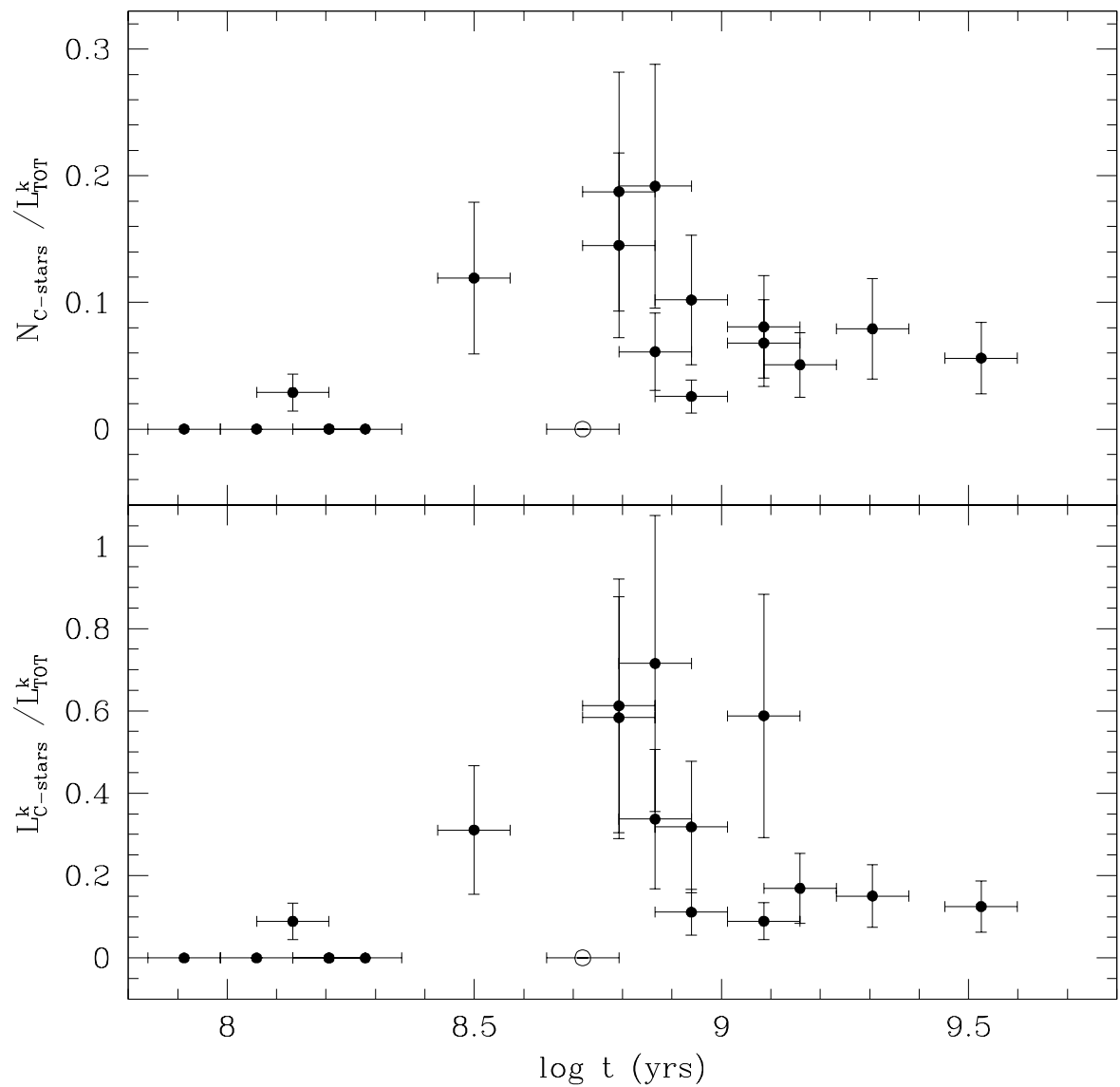}
\caption{{\it Upper panel}: the number of C-stars normalized to the K-band luminosity of the
 cluster as a function of age. {\it Lower panel}: the K-band luminosity of the C-stars 
normalized to the total cluster luminosity as a function of age.
The {\it open circles} marks the intrinsically poor populated cluster NGC~2249.} 
\label{carbon}
\end{figure}

\begin{figure}
\plotone{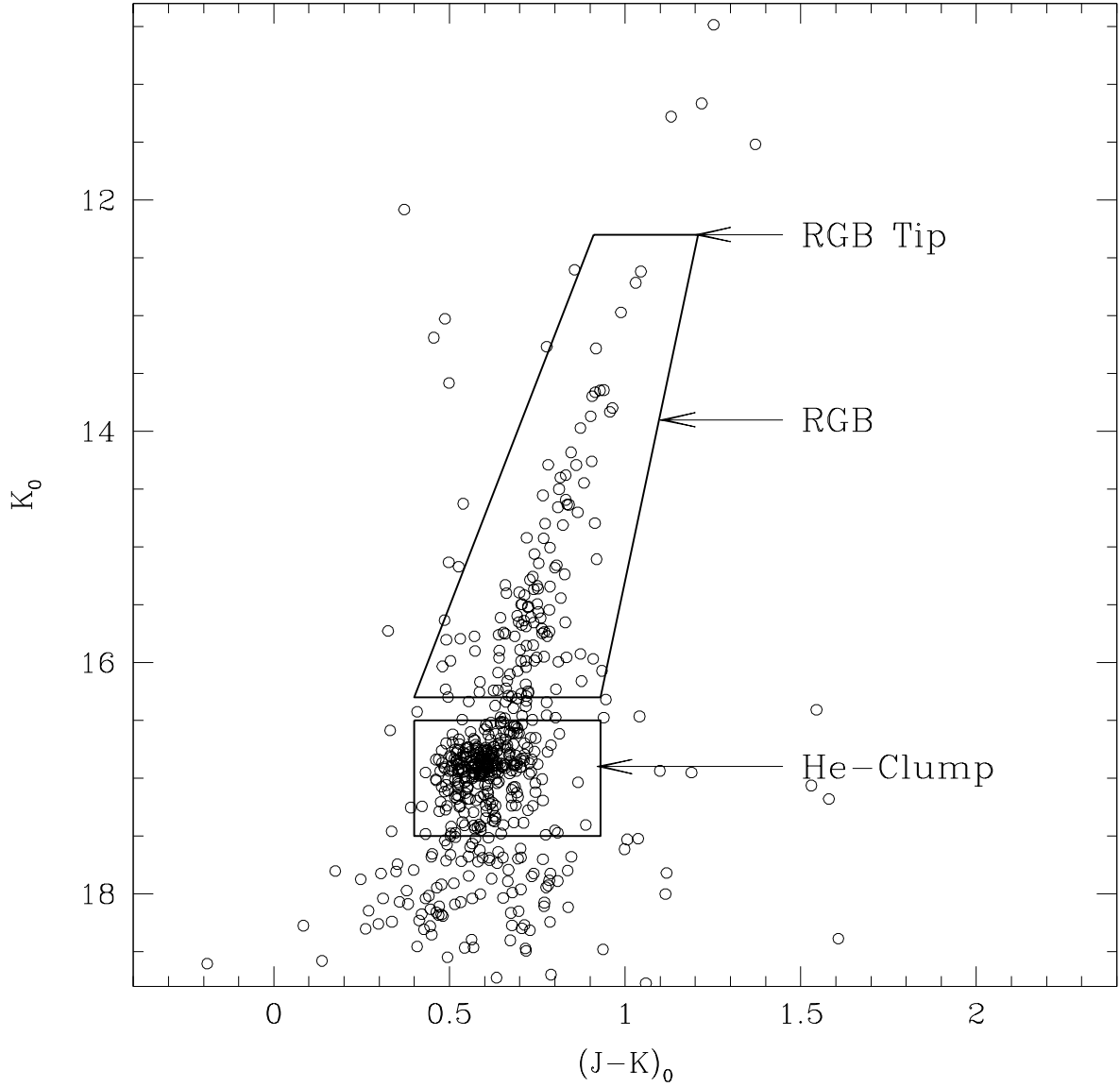}
\caption{An example of a $K_0$, $(J-K)_0$ de-reddened CMD 
with the selection boxes adopted  to distinguish 
the RGB and the He-Clump populations for clusters with $s>39$ (see Fig.~\ref{cmdc2}). 
The position of the RGB Tip is also indicated.}
\label{boxrgb}
\end{figure}

\begin{figure}
\plotone{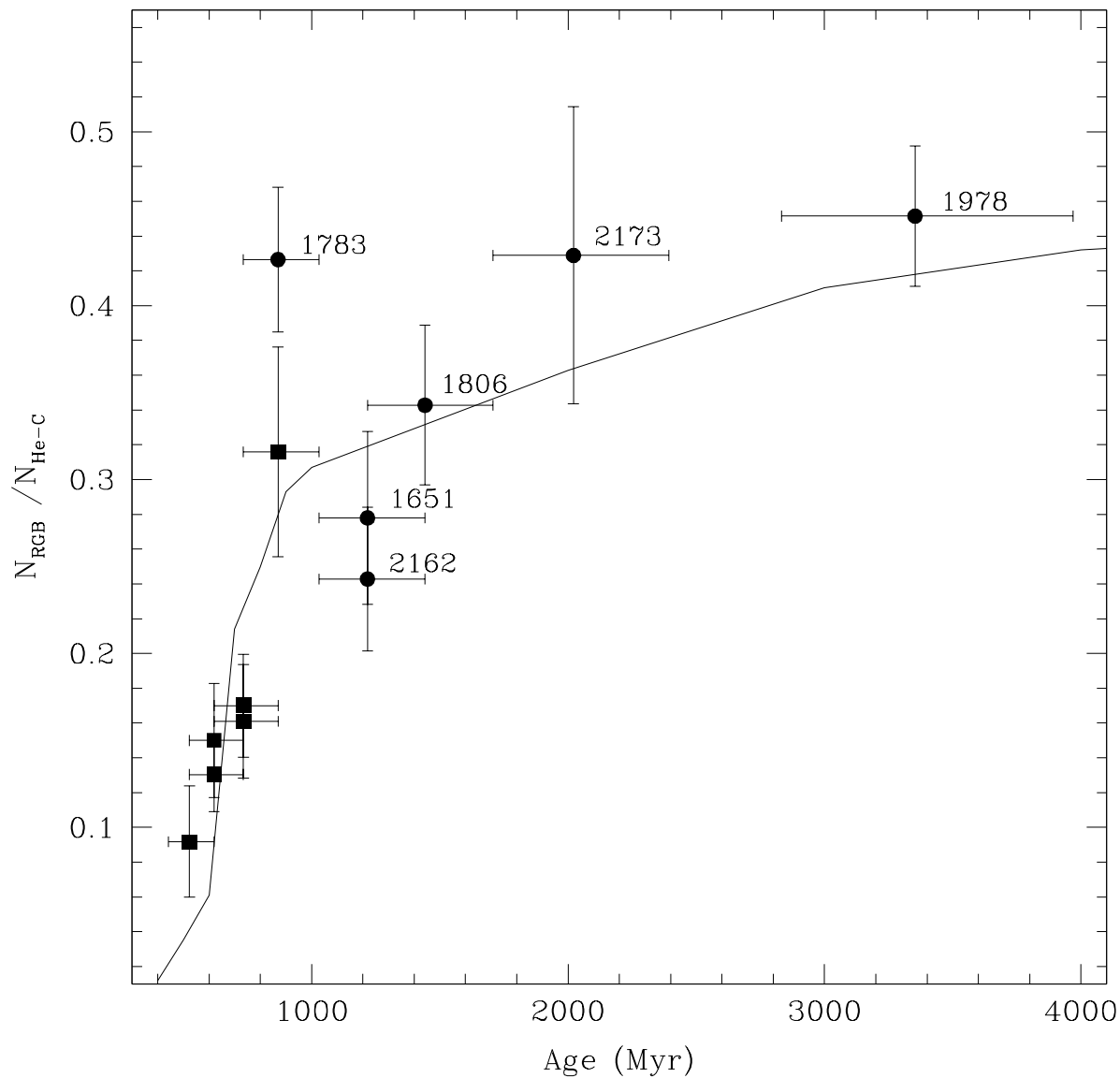}
\caption{Ratio between the number of the bright RGB and He-Clump stars as a function 
of age for the 12 clusters with $s>33$: {\it filled circles} are  
the 6 clusters with $s=39-45$ presented in this paper (see Fig.~\ref{cmdc2}),
{\it filled squares} are the 6 clusters from F04.
Stars belonging to the two populations 
are selected accordingly to the selection boxes shown in Fig.~\ref{boxrgb}. The solid line represents 
the prediction of the canonical theoretical model with
[Z/H]$=-0.33$ \citep{m98}.  }
\label{contrgb}
\end{figure}

\begin{figure}
\plotone{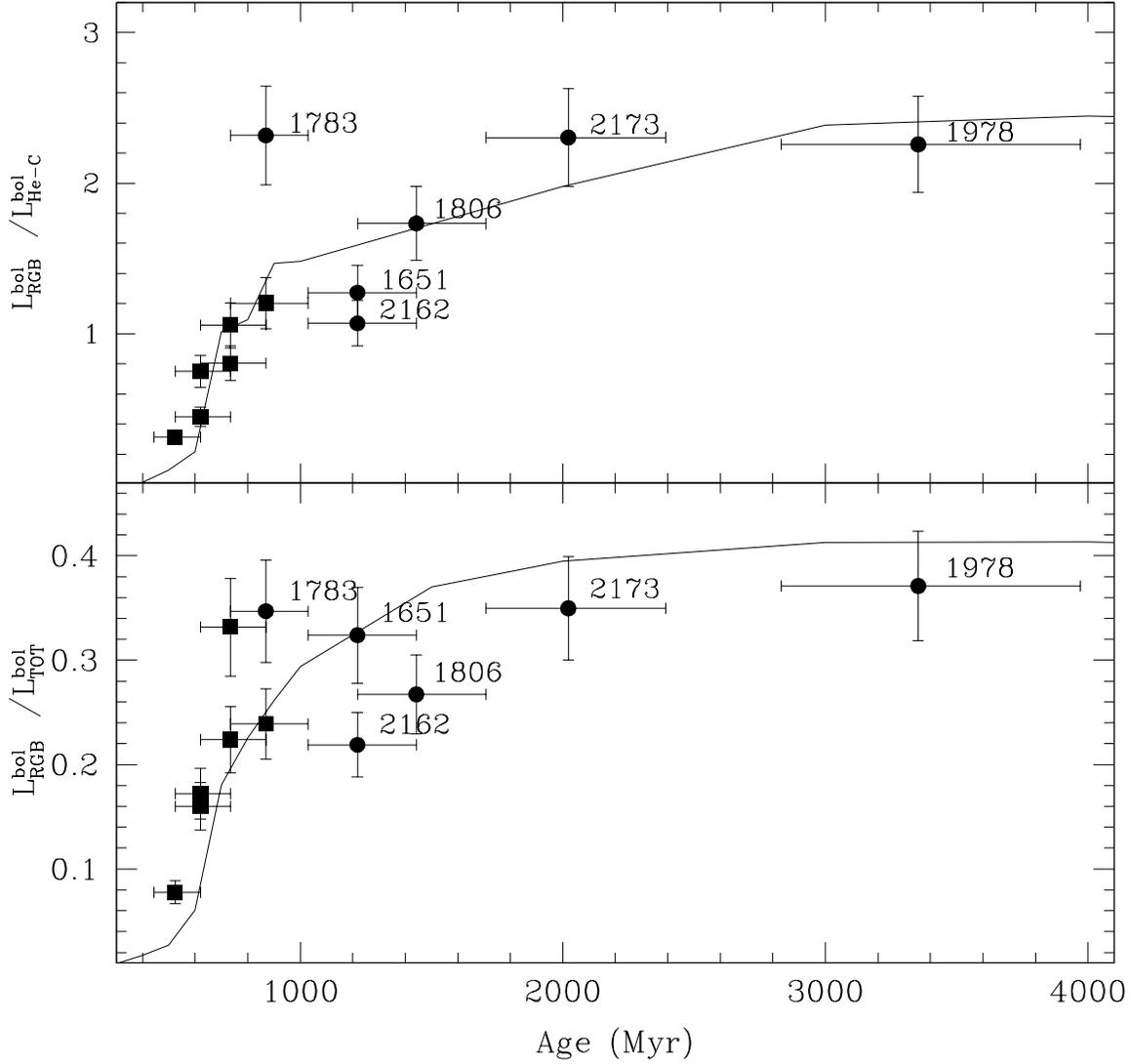}
\caption{ Upper panel:
the bolometric luminosity of the  RGB normalized to the He-Clump 
as a function of age for the 12 clusters with $s>33$.
Symbols and lines are as in Fig.\ref{contrgb}. The line represents 
the prediction of the canonical theoretical model with
[Z/H]$=-0.33$ \citep{m98}. 
Lower panel: the bolometric luminosity of the RGB normalized to the 
total bolometric luminosity for the 12 clusters. The line represents 
the prediction of the canonical theoretical model with
[Z/H]$=-0.33$ \citep{m05}.
}
\label{lumrgb} 
\end{figure}

The behavior  of the quantitative contribution
to the total cluster luminosity
of both the RGB and AGB in terms of age has been
investigated and turns out to be nicely 
in agreement with theoretical prediction of 
canonical models.     
The AGB contribution  to the total cluster luminosity 
starts to be significant 
at $\approx 200~Myr$ and reaches its
maximum peak  at $5-600~Myr$ (see
Fig.~\ref{agbkbol}),  when the RGB Ph-T is starting (see F04
and Figs.~\ref{contrgb} and  \ref{lumrgb}).  Both events
are sharp and last a few hundreds $Myr$ only.

The epoch and duration of the AGB phase-transition estimated from
   our data confirm the semi-empirical modelling of \citet{m98,m05}.
   The latter was based on the calibration of the AGB-fuel
   consumption of canonical tracks, and its partition between C and
   O-type stars, with previous MC GCs data (FMB90, F95).
Furthermore, the epoch and duration of the RGB phase transition derived from our
data agree very well with the theoretical predictions of simple
stellar population models based on canonical tracks and the
fuel-consumption approach \citep{m05}.

This study represents a further step towards  
the calibration of the integrated properties of SSPs 
as a powerful diagnostics for dating complex stellar populations like galaxies.

\acknowledgments
The financial support by  
 the Ministero dell'Istru\-zio\-ne, Universit\`a e Ricerca (MIUR)
is kindly acknowledged.

\clearpage
 
\begin{deluxetable}{lccccccc}
\tablecolumns{8} 
\tablewidth{0pc} 
\tablecaption{Main parameters of the entire sample of observed LMC 
clusters. \label{tabpar}}
\tablehead{ 
\colhead{Cluster}   &  
\colhead{$\alpha $ (J2000)} &   
\colhead{$\delta $ (J2000)} & 
\colhead{\emph{s}} &
\colhead{age ($Myr$)}&
\colhead{[Fe/H]}   & 
\colhead{E(B--V)}   &
\colhead{}} 
\startdata
NGC~2164 &    05:58:55.65   &  -68:31:00.75   & 23 & 81  & $-0.60^{a}$ & 0.10  & this paper \\
NGC~2157 &    05:57:36.74   &  -69:11:53.58   & 25 & 114 & $-0.60^{a}$ & 0.10  &  this paper \\
NGC~2136 &    05:52:58.54   &  -69:29:32.32   & 26 & 135 & $-0.55^{b}$ & 0.10  &  this paper \\
NGC~2031 &    05:33:39.00   &  -70:59:14.54   & 27 & 160 & $-0.52^{b}$ & 0.18  & this paper  \\
NGC~1866 &    05:13:38.88   &  -65:27:53.30   & 27 & 160 & $-0.50^{d}$ & 0.10  &  this paper \\
NGC~2134 &    05:51:57.54   &  -71:05:51.63   & 28 & 190 & $-1.00^{a}$ & 0.10  & this paper  \\
NGC~1831 &    05:06:16.47   &  -64:55:12.76   & 31 & 315 & $+0.01^{c}$ & 0.10  &  this paper \\
NGC~2249 &    06:25:49.50   &  -68:55:14.25   & 34 & 524 & $-0.12^{a}$ & 0.10  & F04\\
NGC~1987 &    05:27:17.29   &  -70:43:56.78   & 35 & 620 & $-1.00^{a}$ & 0.12  & F04\\
NGC~2209 &    06:08:34.87   &  -73:50:06.46    & 35 & 620 & $-1.20^{a}$ & 0.07  & F04\\
NGC~2108 &    05:43:57.30   &  -69:10:55.93  & 36 & 734 & $-1.20^{a}$ & 0.18  & F04\\
NGC~2190 &    06:01:00.67   &  -74:43:29.10  & 36 & 734 & $-0.12^{c}$ & 0.10  & F04\\
NGC~2231 &    06:20:43.67   &  -67:31:13.05   & 37 & 869 & $-0.67^{c}$ & 0.08  & F04 \\
NGC~1783 &    04:59:08.42   &  -65:59:12.75   & 37 & 869 & $-0.45^{a}$ & 0.10  &  this paper \\
NGC~1651 &    04:37:33.86   &  -70:35:09.24   & 39 & 1218& $-0.37^{c}$ & 0.10  &  this paper \\
NGC~2162 &    06:00:30.20   &  -63:43:15.27   & 39 & 1218& $-0.23^{c}$ & 0.07  &  this paper \\
NGC~1806 &    05:02:11.87   &  -67:59:10.11   & 40 & 1442& $-0.23^{c}$ & 0.12  &  this paper \\
NGC~2173 &    05:57:59.28   &  -72:58:42.83   & 42 & 2021& $-0.24^{c}$ & 0.07  &  this paper \\
NGC~1978 &    05:28:45.34   &  -66:14:09.12   & 45 & 3353& $-0.96^{d}$ & 0.10  &  this paper \\
\enddata 		    
\tablecomments{$~~~~$\\
The $s$-parameter is from 
\citet{ef85} and  \citet{girardi95}. Metallicity are from (a) \citet{sp89}, (b) \citet{dirsch00},  
(c) \citet{ols91}, and (d) \citet{hill00}.
Reddening is from \citet{perss83}.}
\end{deluxetable} 

\clearpage

\begin{deluxetable}{lrrrrrr}
\tablecolumns{7}
\tablewidth{0pc}
\tablecaption{Integrated K-magnitude, colors and luminosities of the target clusters.
\label{tabmag}}
\tablehead{
\colhead{Cluster}   &
\colhead{\emph{s}} &
\colhead{K} &
\colhead{J--K} &
\colhead{H--K}&
\colhead{$L_{TOT}^{K}$}&
\colhead{$L_{TOT}^{Bol}$}
}
\startdata
 NGC~2164     & 23    &   8.575  & 0.606  & 0.122       &  19.76   &   12.82       \\
 NGC~2157     & 25    &   8.286  & 0.658  & 0.195       &  25.79   &   15.18       \\
 NGC~2136     & 26    &   7.968  & 0.674  & 0.125       &  34.57   &   19.78        \\
 NGC~2031     & 27    &   8.219  & 0.936  & 0.323       &  28.14   &   11.36       \\
 NGC~1866     & 27    &   7.282  & 0.743  & 0.155       &  65.02   &   33.13       \\
 NGC~2134     & 28    &   9.161  & 0.742  & 0.187       &  11.52   &    5.88    \\
 NGC~1831     & 31    &   8.313  & 0.802  & 0.231       &  25.16   &   11.67        \\
 NGC~2249     & 34    &   9.983  & 0.934  & 0.342       &  5.40    &    2.06       \\
 NGC~1987     & 35    &   8.811  & 0.988  & 0.316       &  16.00   &    5.75       \\
 NGC~2209     & 35    &   8.955  & 1.224  & 0.345       &  13.79   &    3.54       \\
 NGC~2108     & 36    &   8.807  & 1.148  & 0.403       &  16.38   &    4.96       \\
 NGC~2190     & 36    &   9.269  & 1.193  & 0.321       &  10.43   &    2.83       \\
 NGC~2231     & 37    &   9.329  & 1.096  & 0.328       &  9.81   &     2.97       \\
 NGC~1783     & 37    &   7.091  & 1.035  & 0.282       &  77.53   &   25.77        \\
 NGC~1651     & 39    &   8.895  & 0.973  & 0.301       &  14.72   &    5.32       \\
 NGC~2162     & 39    &   9.071  & 1.253  & 0.372       &   12.40   &   3.07       \\
 NGC~1806     & 40    &   7.076  & 1.055  & 0.271       &  79.11   &   25.97        \\
 NGC~2173     & 42    &   9.05   & 1.033  & 0.297       &  12.33   &    4.02       \\
 NGC~1978     & 45    &   7.185  & 0.894  & 0.288       &   71.10   &  28.69        \\
\enddata
\tablecomments{$~~~~$\\
Integrated magnitudes are decontaminated for the field contribution.
K-band and bolometric total luminosities are in units of $10^{4}L_{\odot}$.}
\end{deluxetable}

\clearpage 

\begin{deluxetable}{lcccccccccc} 
\tablecolumns{10} 
\tablewidth{0pc}  
\tablecaption{Star counts and luminosities for AGB and C-stars
\label{tabagb}} 
\tablehead{ 
\colhead{Cluster}  &
\colhead{\emph{s}} &    
\colhead{$N_{AGB}^{Obs}$} & 
\colhead{$N_{AGB}^{Field}$} & 
\colhead{$N_{AGB}^{Dec}$} &
\colhead{$\frac{L_{AGB}^{K}}{L_{TOT}^{K}}$} &  
\colhead{$N_{C-star}$} &  
\colhead{$\frac{N_{C-star}}{L_{TOT}^{Bol}}$} & 
\colhead{$\frac{N_{C-star}}{L_{TOT}^{K}}$} & 
\colhead{$\frac{L_{C-star}^{K}}{L_{TOT}^{K}}$}
}
\startdata 
 NGC~2164   & 23 &     2    & 0  &	2  & 0.10  & 0 &	 0 &	     0  &  0	  \\
 NGC~2157   & 25 &     9    & 1  &	8  & 0.52  & 0 &	 0 &	     0  &  0	  \\
 NGC~2136   & 26 &     9    & 1  &	8  & 0.31  & 1 &     0.05 &	 0.03   &  0.09     \\
 NGC~2031   & 27 &     7    & 1  &	6  & 0.64  & 0 & 	 0 &	     0  &  0	  \\
 NGC~1866   & 27 &    12    & 0  &     12  & 0.36  & 0 &	 0 &	     0  &  0	  \\
 NGC~2134   & 28 &     2    & 1  &	1  & 0.71  & 0 &	 0 &	     0  &  0	  \\
 NGC~1831   & 31 &     7    & 1  &	6  & 0.45  & 3 &     0.26 &	 0.12   &  0.31     \\
 NGC~2249   & 34 &     1    & 0  &	1  & 0.32  & 0 &	 0 &	     0  &  0     \\
 NGC~1987   & 35 &     9    & 4  &	5  & 0.90  & 3 &     0.52 &	 0.19   &  0.61     \\
 NGC~2209   & 35 &     4    & 0  &	4  & 0.72  & 2 &     0.56 &	 0.14   &  0.58    \\
 NGC~2108   & 36 &     5    & 1  &	4  & 0.87  & 1 &     0.20 &	 0.06   &  0.34    \\
 NGC~2190   & 36 &     2    & 0  &	2  & 0.74  & 2 &     0.71 &	 0.19   &  0.72    \\
 NGC~2231   & 37 &     1    & 0  &	1  & 0.32  & 1 &     0.34 &	 0.10   &  0.32    \\
 NGC~1783   & 37 &    16    & 1  &     15  & 0.36  & 2 &     0.08 &	 0.03   &  0.11   \\
 NGC~1651   & 39 &     4    & 0  &	4  & 0.51  & 1 &     0.19 &	 0.07   &  0.09    \\
 NGC~2162   & 39 &     4    & 1  &	3  & 0.73  & 1 &     0.32 &	 0.08   &  0.59    \\
 NGC~1806   & 40 &    13    & 4  &	9  & 0.22  & 4 &     0.15 &	 0.05   &  0.17     \\
 NGC~2173   & 42 &     5    & 1  &	4  & 0.53  & 1 &     0.25 &	 0.08   &  0.15    \\
 NGC~1978   & 45 &    13    & 1  &     12  & 0.25  & 4 &     0.14 &	 0.06   &  0.13    \\	   
\enddata 
\tablecomments{$~~~~$\\
Star counts  
are corrected for incompleteness.
K-band and bolometric luminosities are in units of $\rm 10^4~L_{\odot}$.}
\end{deluxetable} 

\clearpage 

\begin{deluxetable}{lcccccc}
\tablecolumns{7} 
\tablewidth{0pc} 
\tablecaption{Star counts and bolometric luminosities for RGB and He-Clump stars.
\label{tabrgb}} 
\tablehead{ 
\colhead{Cluster} &  
\colhead{\emph{s}} &
\colhead{$N_{RGB}~\rm ^{(a)}$} &   
\colhead{$N_{He-C}~\rm ^{(a)}$} & 
\colhead{$L_{RGB}^{bol~~\rm (b)}$} &
\colhead{$L_{He-C}^{bol~~~\rm (b)}$} &
\colhead{}
}
\startdata
NGC~2249  & 34  &   9 &  98 & 0.16 & 0.51    & F04\\	    
NGC~1987  & 35  &  42 & 322 & 0.92 & 2.05  & F04\\	    
NGC~2209  & 35  &  24 & 160 & 0.61 & 0.81  & F04\\	    
NGC~2108  & 36  &  40 & 231 & 1.11 & 1.38   & F04\\	    
NGC~2190  & 36  &  28 & 174 & 0.94 & 0.89    & F04\\	    
NGC~2231  & 37  &  36 & 114 & 0.71 & 0.59   & F04 \\	    
NGC~1783$^{(c)}$  & 37  & 150 & 352 & 4.58 & 1.98  &  this paper \\ 
NGC~1651  & 39  &  43 & 177 & 1.16 & 1.09  &  this paper \\ 
NGC~2162  & 39  &  40 & 143 & 0.99 & 0.78  &  this paper \\ 
NGC~1806$^{(c)}$  & 40  &  75 & 218 & 2.25 & 1.30  &  this paper \\ 
NGC~2173  & 42  &  36 &  84 & 1.11 & 0.48  &  this paper \\ 
NGC~1978$^{(c)}$  & 45  & 182 & 402 & 5.03 & 2.23  &  this paper \\ 
\enddata
\tablecomments{$~~~~$\\
$\rm (a)$~Star counts  
are corrected for incompleteness and field contamination.\\
$\rm (b)$~Bolometric luminosities are in units of $\rm 10^4~L_{\odot}$.\\
$\rm (c)$~Due to severe crowding conditions, star counts and luminosities have 
been computed only in the outer B and C regions, see Fig.\ref{map}.}
\end{deluxetable} 

\clearpage 

\end{document}